\documentclass[preprint,aps,epsf]{revtex4}
\usepackage{graphicx}
\newcommand{\beq}{\begin{equation}}
\newcommand{\eeq}{\end{equation}}
\newcommand{\beqa}{\begin{eqnarray}}
\newcommand{\eeqa}{\end{eqnarray}}
\newcommand{\beqar}{\begin{eqnarray*}}
\newcommand{\eeqar}{\end{eqnarray*}}

\def \d {{\delta r}}

\def \O {{\bf O}}
\def \H {{\bf H}}

\def \T {{\bf T_A}}
\def \la {\langle}
\def \ra {\rangle}

\def \x {{\bf x}}
\def \y {{\bf y}}
\def \p {{\bf p}}

\def \px {{\bf P_x}}
\def \py {{\bf P_y}}

\def \e {\epsilon}
\def\figloc#1#2#3
{
\begin{figure}[ht]
      \includegraphics[scale=0.4]{fig#1.ps}
\caption{#3
}
\end{figure}
}

\def\figloccent#1#2#3
{
\begin{figure}[ht]
      \includegraphics[scale=0.4]{fig#1.ps}
\caption{#3
}
\end{figure}
}
\def\figlocbig#1#2#3
{
\begin{figure}[ht]
      \includegraphics[scale=0.6]{fig#1.ps}
\caption{#3
}
\end{figure}
}
\begin{document}
%\listoffigures
%\epsfverbosetrue
%@twocolumnfalse\endcsname

%\input epsf

\title{ \bf\Large
Temporal Ordering in Quantum Mechanics
}

\author{ {
J. Oppenheim$^{(a),(b)}$,
%\footnote{\it jono@phys.ualberta.ca}
B. Reznik$^{(c)}$,
%\footnote{\it reznik@post.tau.ac.il}
and W. G. Unruh$^{(a)}$}
%\footnote{\it unruh@physics.ubc.ca\\}
{\ } \\
(a) {\it \small   Department of Physics,  University of British
Columbia,
6224 Agricultural Rd. Vancouver, B.C., Canada
V6T 1Z1}\\
(b) {\it \small Department of Physics, University of Alberta, 412 Avadh Bhatia
Physics Laboratory, Alta., Canada, T6G 2J1}  \\
(c) {\it \small School of Physics and Astronomy, Tel-Aviv University, Tel-Aviv
  69978,  Israel} }
%\maketitle

\begin{abstract}

We examine the measurability of the temporal ordering of two events, as
well as event coincidences.
In classical mechanics, a measurement of the order-of-arrival
of two particles is shown to be
equivalent to a measurement
involving only one particle (in higher dimensions).
In quantum mechanics, we find that diffraction effects introduce
a minimum inaccuracy to which the temporal order-of-arrival can
be determined unambiguously.
The minimum inaccuracy of the measurement is
given by $\delta t =\hbar/ {\bar E}$ where ${\bar E}$ is
the total kinetic energy of the two particles.
Similar restrictions apply to the case of coincidence measurements.
We show that these limitations are much weaker than limitations on
measuring the time-of-arrival of a particle to a fixed location.
%It is shown that in general one cannot prepare a
%two particle state where the two particles always arrive within a time of
%$1/{\bar E}$ of each other.

\end{abstract}
%\vskip2pc]
%\vskip2cm
\maketitle
%\newpage
%
%
%
\section{Introduction}
%\noindent
In quantum mechanics, one typically measures operators at fixed times $t$.
For example, one can measure the position of a particle at any given time,
and obtain a precise result.  One could also consider the "dual" situation
in which one tries to measure at what time a particle arrives to a fixed
location $x_a$.  This problem of time-of-arrival \cite{allcock}
has been extensively discussed in the literature \cite{muga}.

Although the time $t$ is a well defined parameter in the Schr\"odinger
equation, Pauli has shown that it cannot correspond to an operator
for systems which have an energy bounded from below \cite{pauli}.
Likewise, for general Hamiltonians, there is no operator which corresponds to
the time of an event such as the time-of-arrival of a particle to a fixed
location \cite{aharonov}.
In addition, if one wishes to operationally measure the time-of-arrival
by coupling the system to a clock, then one finds that
one cannot measure the time-of-arrival to an accuracy
better than $\hbar/{\bar E_k}$ where ${\bar E_k}$ is the kinetic energy of the
particle\cite{allcock,aharonov,tmeas}. The limitation is
based on calculations from a wide variety of different measurement models, as
well as general considerations, however, there is no known proof of this
result.
%%%%%%%%%%%%

There have been attempts to circumvent these difficulties
\cite{rovelli}\cite{circ}\cite{muga}, usually involving a modified
time-of-arrival operator or POVM measurements. Such operators can be measured
``impulsively'' by interacting with the system at a certain (arbitrary) instant
of time. In this manner, one can attempt to measure the time-of-arrival even
though the particle has not arrived (and in fact, may never arrive, regardless
of what the result of the time-of-arrival measurement yields)\cite{tmeas}.
These procedures, are
hence conceptually and operationally very different from
the case of continuous measurements discussed here.

%%%%%%%%%%%%
One can also ask, given two events $A$ and $B$, whether one can measure
which event occurred first.  Surprisingly,
there does not appear to be any discussion of this in the literature,
even though we believe it is a much more primitive and fundamental concept.
In this paper, we are interested in whether the well defined classical
concepts of temporal ordering have a quantum analogue.  In other words,
given two quantum mechanical systems, can we measure which
system attains a particular state first.  Can we decide whether an event
occurs in the past or future of another event.

Classically, one can couple the system to a device which is triggered
when an event
occurs, and records which event happened first.  One can consider a similar
measurement scheme in quantum mechanics which classically would correspond to a
measurement of order of events.  One can then ask whether such a quantum
measurement scheme is possible.

The fact that there is a limitation to measurements of the time of an event
leads one to suspect that the ordering of events may not be an
unambiguous concept in quantum mechanics.
However, for a single quantum event $A$, although one cannot determine the
time an event occurred to arbitrary accuracy, it can be argued that one
can often measure  whether $A$ occurred before or after a fixed time $t_B$ to
any desired precision.

Consider a quantum system initially prepared in a state $\psi(0)$
and an event $A$ which corresponds to some projection operator $\Pi_A$
acting on this state.
For example, we could initially prepare an atom in an excited state,
and $\Pi_A$ could represent a projection onto all states where the
atom is in its ground state i.e. the atom has decayed.  $\psi(0)$
could also represent a particle localized in the region $x<0$ and
$\Pi_A$ could be a projection onto the positive x-axis.  In this case,
the event $A$ corresponds to the particle arriving to $x=0$.

If the state evolves irreversibly to a state for which $\Pi_A \psi(t) =1$,
then we can easily measure whether the event $A$ has occurred at any time
$t$.  We could therefore measure whether a free particle
arrives to a given location before or after a classical time $t_B$.
Of course, for many systems, the system will not irreversibly evolve to the
required state.  For example, a particle influenced by a potential may cross
over the origin many times\footnote{Here, and throughout this paper, we will
sometimes use language which refers to objective facts about a particle's
motion.  It should be understood that these descriptions refer to the results
of measurements made on these particles.  For example, it can be measured
that a particle is traveling towards the origin in the case where one can make
a weak measurement of position and momentum.}.  However, for an
event such as atomic decay, the probability of the atom being re-excited is
relatively small, and one can argue that the event is effectively irreversible.

For the case of a free particle which has been measured to be traveling towards
the origin from $x<0$ one can argue that if at a later time we measure the
projection operator onto the positive axis and find it there, then the particle
must have arrived to the origin at some earlier time.  This is in some sense a
definition, because we know of no way to measure the particle being at the
origin without altering its
evolution (or being extremely lucky and happening to measure the particle's
location when it is at the origin).

While measuring whether an event happened before or after a fixed
time $t_B$ may be possible, we will find
that for two quantum events, one cannot in general measure whether the time
$t_A$ of event $A$, occurred before or after the time $t_B$ of event $B$.

In Section \ref{or:whofirst}, confining ourselves to a particular example of
order of events, we will consider the question of order of
arrival in quantum mechanics.
Given two particles, can we determine which particle
arrived first to the location $x_a$.
Using a model detector, we find that there is always
an inherent inaccuracy in this type of measurement given by $\hbar/{\bar E}$ where
${\bar E}$ is the typical total energy of the two particles.  This seems to
suggest that the notion of past and future is not a well defined observable in
quantum mechanics.

We will see that this inaccuracy limitation on the measurement of
order-of-arrival is weaker than the inaccuracy on measurements of
time-of-arrival.  If one attempted to measure the
order-of-arrival by measuring the time-of-arrival of both particles,
then the limitation on the measurement accuracy is much greater, being
$\hbar/min\{E_x, E_y
\}$ where $E_x$ and $E_y$ are the typical energies of each individual particle.

In the present article we will consider only
continuous measurements in which the detector is left ``open''
for a long duration.
One can also formally define
an order-of-arrival operator like
\beq
\O=sgn(\T_x - \T_y)
\eeq
where $T_x$ and $T_y$ are the time-of-arrival operators
\beq
\T=\frac{mx_a}{\p}-m\frac{1}{\sqrt{\p}}\x\frac{1}{\sqrt{\p}} \,\,\,\,\, .
\eeq

As already noted, if one measures such an operator
%Such operators can be measured
%``impulsively'' by interacting with the system at a certain (arbitrary) instant
%of time. In this manner,
one is measuring which event occurred first,
even though neither event has in fact occurred (and
may not occur).
The measurement of an operator, and the continuous, "operational"
methods discussed here, are therefore rather different.
%hence conceptually and operationally very different from
%the case of continuous measurements discussed here.
Furthermore, the time-of-arrival operator cannot be
self-adjoint \cite{aharonov},
and therefore has complex eigenvalues and eigenstates \cite{reedsimon}.
However, it can be modified\cite{rovelli}. We believe that modifying
the operator
causes several technical as well as fundamental difficulties. For example, it
has been shown \cite{oppenheim}, that the eigenstates of
modified time-of-arrival operators such as those in \cite{rovelli}
no longer describe events of arrival at a definite time.
We anticipate similar difficulties for the case of the order-of-arrival
operator.

In Section \ref{or:coincidence} we discuss measurements of coincidence.  I.e.,
can we determine whether both particles arrived at the same time.  Such
measurements allow us to change the accuracy of the device before each
experiment.  We  find that the measurement fails when the accuracy is made
better than $\hbar/{\bar E}$.
%Limitations
%on the accuracy of coincidence measurements are more accesable to real
%experiments since the coincident arrival of two particles, rather than the
%time-of-arrival  of a single particles is more accurately measurable in the
%labratory.

In Section \ref{or:micro} we discuss the relationship between ordering of
events and the resolving power of ``Heisenberg's microscope``\cite{heisenberg},
and argue that in general, one cannot prepare a two particle state which is
always coincident to within a time of $\hbar/{\bar E}$.
In the following we use units such that $\hbar=1$.

\section{Which first? }\label{or:whofirst}

Consider two free particles (which we
will label as x and y) initially
localized to the right of the origin, and traveling to the left.  We then
ask whether one can measure which particle arrives to the origin first.
The Hamiltonian for the system and measuring apparatus is given by
\beq
\H=\frac{\px^2}{2m_1} + \frac{\py^2}{2m_2} + \H_i \label{eq:genham}
\eeq
where $\H_i$ is some interaction Hamiltonian which is used to perform the
measurement.  One possible choice for an interaction Hamiltonian is
\beq
\H_i=\alpha \delta (\x)\theta (-\y) \label{eq:knifeham}
\eeq
with $\alpha$ going to infinity.

If the y-particle arrives before the x-particle, then the x-particle will be
reflected back.  If the y-particle arrives after the x-particle,
then neither particle sees the potential, and both particles will continue traveling
past the origin.  One can therefore wait a sufficiently long
period of time, and measure where the two particles are.  If both the
x and y particles are found past the origin, then we know that the
x-particle arrived first.  If the y-particle is found past the origin
while the x-particle has been reflected back into the positive x-axis
then we know that the y-particle arrived first.

Classically, this method would appear to unambiguously measure which
of the two particle arrived first.  However, in quantum mechanics,
this method fails.  From (\ref{eq:genham}) we can see that the problem of
measuring which particle arrives first is equivalent to deciding
where a single particle traveling in a plane arrives.  Two particles
localized to  the right of the origin is
equivalent to a single particle localized in the first quadrant (see
Figure 1).  The question of which particle arrives first, becomes equivalent
to the question of whether the particle crosses the positive x-axis or the
positive y-axis.
%
%\begin{figure}[h]
%\begin{center}
%\leavevmode
%\centering
%the begin and end figure is for a figure section, and includes a caption
%\vspace {3cm}
%\hspace {4.3cm}
%\epsfysize=3.0in
%\epsfbox[-230 132 575 700]{essai2.ps}
% this is another way of doing it
%\epsfbox{knife.ps}
%\caption[Order of arrival detector]{ caption}
%%$\protect\sqrt{\frac{\Delta}{m}}
\figloc{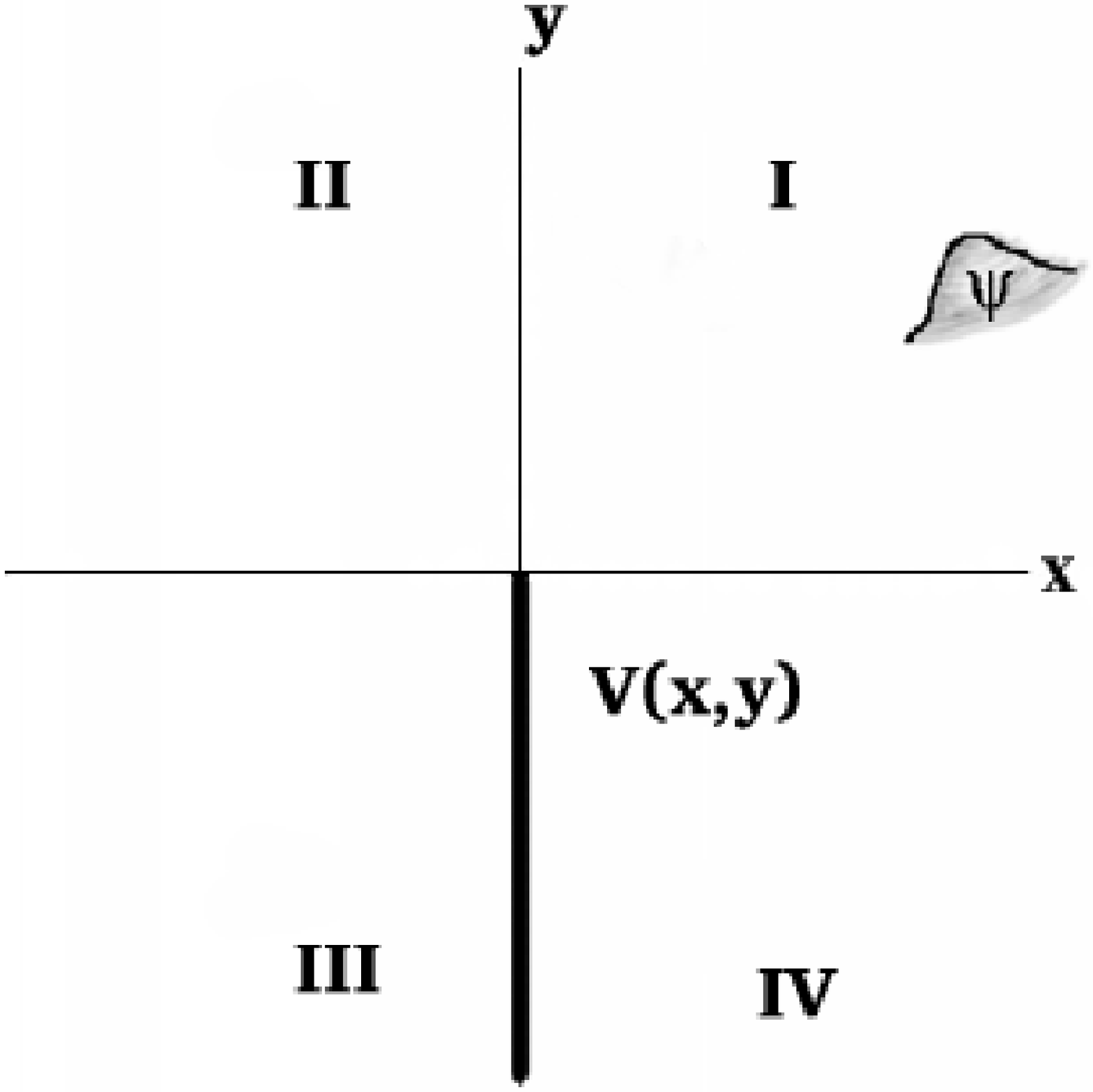}{}{The measurement of order-of-arrival for two particles in one dimension, is
equivalent to scattering of a straight edge of one particle in
two-dimensions. }
%this is fig \ref{fig:knife}
%\end{center}
%\end{figure}
%
%
%\epsfysize=3in
%    \centerline{\epsfbox{fig1.ps }
%\addcontentsline{lof}{figure}{test}}
%    \centerline{Figure 1}
%    {\raggedright\it   caption text }
%    \bigskip
%

The equivalence between the two-particle system and the single particle
system in higher dimensions can be seen by performing the
canonical transformation \beqa
\px &\longrightarrow &  \sqrt{m_1\over M}\px, \ \ \ \
\py \longrightarrow  \sqrt{m_2\over M}\py \nonumber \\
\x &\longrightarrow & \sqrt{M\over m_1}\x, \ \ \ \ \ \
\y \longrightarrow  \sqrt{M\over m_2}\y \label{eq:remap}
\eeqa
and rescaling $\alpha \to \sqrt{M/m_1}\alpha$.  Our Hamiltonian now looks like
that of a single particle of mass $M$
scattering off a thin edge in two dimensions.
Classically, the event $x$ arriving first,  corresponds to the  case that
the particle does not
scatter off the edge and  travels to quadrant III.
The event of $y$ arriving first corresponds to scattering
 off the edge to quadrant IV.

However, quantum mechanically, we find that sometimes the particle
is found in the two classically forbidden regions, I and II.
If the particle is found in either of these two regions, then we
cannot determine which particle arrived first.

The solution for a plane wave which makes an angle $\theta_o$ with the
x-axis is well known\cite{morseandf}.  If the boundary condition
is such that $\psi(r,\theta)=0$ on the negative y-axis, then the solution
is
\beq
\psi(r,\theta) = \frac{1}{\sqrt{i\pi}}\left\{
e^{-ikr\cos(\theta-\theta_o)}\Phi[\sqrt{2kr}\cos(\frac{\theta-\theta_o}{2})]
-
e^{ikr\cos(\theta+\theta_o)}\Phi[-\sqrt{2kr}\sin(\frac{\theta+\theta_o}{2})]
\right\}
\eeq
where $\Phi(z)$ is the error function.

Asymptotically, this solution looks like
\beq
\psi  \simeq \left\{
\begin{array}{ll}
e^{-ikr\cos(\theta-\theta_o)}
+
f(\theta) \frac{e^{ikr}}{\sqrt{r}} \,\,\,\,\,\,
&
-\theta_o < \theta < \pi + \theta
\\
e^{-ikr\cos(\theta-\theta_o)} -
e^{ik r \cos(\theta+\theta_o)}
+
f(\theta)\frac{e^{ikr}}{\sqrt{r}} \,\,\,\,\,\,
&
- \theta_o > \theta > -\pi/2
\\
f(\theta)\frac{e^{ikr}}{\sqrt{r}} \,\,\,\,\,\,
&
\pi-\theta_o < \theta < 3\pi/2
 \end{array}
\right.
\eeq
where
\beq
f(\theta) = -\sqrt{\frac{i}{8\pi k}}
\left[ \frac{1}{\sin(\frac{\theta+\theta_o}{2})}+
\frac{1}{\cos(\frac{\theta-\theta_o}{2})} \right] \,\, .
\eeq
The above approximation is not valid when $\cos(\frac{\theta-\theta_o}{2})$ or
$\sin(\frac{\theta+\theta_o}{2})$ is close to zero.

Since we demanded that the particle was initially localized in the first
quadrant, the initial wave cannot be an exact plane wave, but we can
imagine that it is a plane wave to a good approximation.

We see from the solution above that the particle can be found in
the classically forbidden regions of quadrant I and II.
For these cases, we cannot determine which particle arrived first.
This is due to interference which occurs when the particle is
close to the origin (the sharp edge of the potential).
The amplitude for being scattered off the region around the edge in
the direction $\theta$ is given
by $|f(r,\theta)|^2$.

It might be argued that since these particles scattered, they must
have scattered off the potential, and therefore they represent
experiments in which the y-particle arrived first.  However, this
would clearly over count the cases where the y-particle arrived first.
We could have just as easily have placed our potential on the negative
x-axis, in which case, we would over-count the cases where the x-particle
arrived first.

In the "interference region" we cannot have confidence that our
measurement worked at all.  We should therefore define a
"failure cross section" given by
\beqa
\sigma_f & =&  \int_0^{2\pi} |f(\theta)|^2 \nonumber\\
&=& \frac{1}{k \cos(\frac{\theta_o}{2}) } \label{eq:failure}
\eeqa

From (\ref{eq:failure}) we can see that cross section for scattering
off the edge is the size of the particle's wavelength
multiplied by some angular dependence.
Therefore, if the particle arrives within a distance of the origin given by
\beq
\delta x > 2/k   \,\,\,\,\,\,
\eeq
the measurement will fail.
We have dropped the angular dependence from
(\ref{eq:failure}) -- the angular dependence is not of physical importance
for measuring which particle came first, as it depends on the details of the
potential (boundary conditions) being used.  The particular potential we have
chosen is not symmetrical in x and y.

From this we can conclude that if the particle arrives to within
one wavelength of the origin, then there is a high probability
that the measurement will fail.

If we want to relate this two-dimensional scattering problem back to
two particles traveling in one dimension, we need to use the relation
\beq
\delta t \simeq \frac{m \delta x}{k}
\eeq
In other words, our measurement procedure relies on making an inference
between time measurements and spatial coordinates.
The last two equations then give us
\beq
\delta t > \frac{1}{E} \,\,\,\,\,\, . \label{eq:ordlim}
\eeq
One will not be able to determine which particle arrived first,
if they arrive within
a time $1/E$ of each other, where $E$ is the total kinetic
energy of both particles.  Note that Equation (\ref{eq:ordlim})
is valid for a plane wave with definite momentum $k$.  For
wave functions for which $dk << k$, one can replace $E$ by
the expectation value $\la E \ra$.  However, for wave functions
which have a large spread in momentum, or which have a number of
distinct peaks in $k$, then to ensure that the measurement almost
always works, one must measure the order of arrival with an accuracy
given by
\beq
\delta t > \frac{1}{{\bar E}}
\eeq
where ${\bar E}$ is the minimum typical total energy
\footnote{For example, one need not be concerned with exponentially small
tails in momentum space, since the contribution of this part of the wave
function to the probability distribution will be small.  If however, $\psi(E)$
has two large peaks at $E_{small}$ and $E_{big}$ spread far apart, then if
$\delta t$ does not satisfy $\delta t > 1/E_{small}$ one will get a distorted
probability distribution.  For a discussion of this, see \cite{aharonov}}.
Hence we conclude that if the particles are
coincident to within $1/{\bar E}$, then the measurement fails.

It is rather interesting that this measurement limitation
is less strict than the one obtained if we were to measure the time-of-arrival
of each particle individually.  This can be seen from the mapping of
Eq. \ref{eq:remap} since the total energy ${\bar E}=E_x+E_y$ where $E_x$ and
$E_y$ are the energies of each individual particle.  The
limitation on measurements of the time-of-arrival of each particle is given
by $1/E_x$ and $1/E_y$ \cite{aharonov}.  Therefore, if we use time
of arrival measurements to determine the order of arrival, the
minimal inaccuracy will have to be $1/min\{E_x,E_y\}$ which can be
considerably worse than $1/(E_x+E_y)$ using the method outlined
above.

The extreme limit, where one of the particles has a very high
energy is then rather interesting.
We have argued in the previous section that for the case of a single
event, we can measure with arbitrary accuracy
if the event occurred before or after a certain {\em
given} time $t_0$.
Indeed, let us consider the above setup in the special limit
that $E_y \gg E_x $ with $E_y \to \infty$.
The diffraction pattern in this case is  completely controlled
by the $y$ particle and
 $\delta t > {1\over \bar E} \sim {1\over E_y} \to 0$.
Furthermore for the case $dy \ll dE_y \ll E_y$, the location $y$
of the energetic particle can serve as a good ``clock''\cite{bennicasher} and
has  a well defined time-of-arrival to $y=0$.
Hence the initial state of the
$y$ particle defines (up to $1/E_y\to 0$)
the time-of-arrival of the y-particle, $t_0=t_A(y=0)$.
The final states of the ``clock'' hence determines whether
 the $x$  particle
arrived before or after $t=t_0$.
If $y_{final}>0$ we conclude that $t_A(x=0)<t_0$ and if
$y_{final}<0$ that $t_A(x=0)>t_0$.

On can create a full clock, by considering many heavy "y" particles,
and determining whether the "x" particle came before or after each one of
them.  Increasing the number of "y" particles and having them arrive
at regularly spaced intervals would then
constitute a measurement of time-of-arrival.  We would then expect to recover
the limitation of reference \cite{aharonov} as the density of
"y" is increased.

\section{Coincidence} \label{or:coincidence}

In the previous model for measuring which particle arrived first, we
found that if the two particles arrived to within $1/{\bar E}$ of each other,
the measurement did not succeed.  The width $1/{\bar E}$ was an
inherent inaccuracy which could not be overcome.  However, in our
simple model, we were not able to adjust the accuracy of the measurement.  

It is therefore instructive to consider a measurement of ``coincidence'' alone
for which one can quite naturally adjust the accuracy of the experiment.
Given two particles traveling towards the origin, we ask whether
they arrive within a time $\delta t_c$ of each other.  If the particles
do not arrive coincidently, then we do not concern ourselves with
which arrived first.  The parameter $\delta t_c$ can be adjusted, depending
on how accurate we want our coincident ``sieve'' to be. 
We will once again find
that one cannot decrease $\delta t_c$ below $1/{\bar E}$ and still have the
measurement succeed.

A simple model for a coincidence measuring device
can be constructed in a manner similar to (\ref{eq:knifeham}).
Mapping the problem of two particles to a single particle in two dimensions, 
we could consider an infinite potential strip of length $2a$ and infinitesimal
thickness, placed at an angle
of $\pi/4$ to the x and y axis in the first quadrant (see Figure 2).
Particles which miss the strip, and travel into the third quadrant
are not coincident, while particles which bounce back off the strip
into the first quadrant are measured to be coincident. 
I.e. if the x-particle is located within a distance $a$ of the origin when
the y-particle arrives (or visa versa), then we call the state coincident. 

%\begin{figure}
%the begin and end figure is for a figure section, and includes a caption 
%\vspace {3cm}
%\hspace {3.3cm}
%\epsfysize=2.0in
%\epsfbox[-230 132 575 700]{essai2.ps}
% this is another way of doing it
%\epsfbox{slab.ps}
%\caption[]{ 
%$\protect\sqrt{\frac{\Delta}{m}}
\figloccent{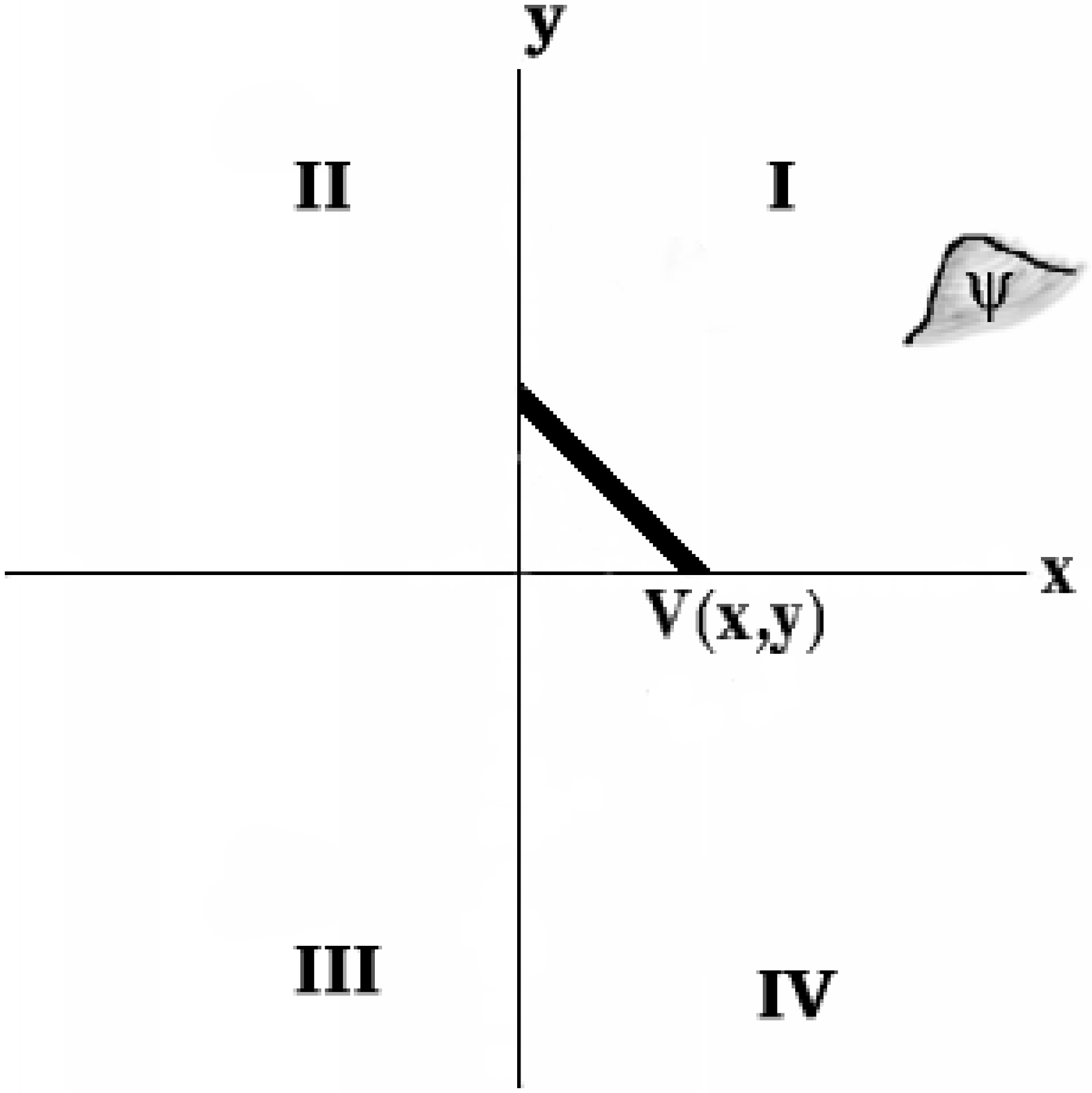}{}{Potential for measuring whether two particles are coincident.
}
%\end{figure}

Classically, one expects there to be
a sharp shadow behind the strip.  Quantum mechanically, we once again
find an interference region around the strip which scatters particles
into the classically forbidden regions of quadrant two and four.  
The shadow is not sharp, and we are not always certain whether the
particles were coincident.

A solution to plane waves scattering off a narrow strip is well known and can
be found in many quantum mechanical texts (see for example \cite{morseandf} where the scattered
wave is written as a sum of products of Hermite polynomials and Mathieu
functions). However, for our purposes, we will find it convenient to consider
a simpler model for measuring coincidence, namely, an infinite circular potential
of radius $a$, centered at the origin.

\beq
H_i=\alpha V(r/a)
\eeq
where $V(x)$ is the unit disk, and we take the limit $\alpha \rightarrow\infty$.

It is well known that if $a<1/k$, then there will not be a well-defined
shadow behind the disk.
To see this, consider a plane wave coming in from negative x-infinity.  It can
be expanded in terms of  the Bessel function $J_m(kr) $ and then written
asymptotically  ($r \gg 1$) as a sum
of incoming and outgoing circular waves.
\beqa
e^{ikx} &=& \sum_{m=0}^{\infty}\epsilon_m i^m J_m(kr)\cos{m\theta} \nonumber\\
& \simeq& \sqrt{\frac{1}{2\pi i kr}} \left[
e^{ikr} \sum_{m=0}^{\infty} \epsilon_m \cos{m\theta} 
+ 
i e^{-ikr} \sum_{m=0}^{\infty} \e_m \cos{m(\theta-\pi)} 
\right]\,\,\, . \label{eq:inout}
\eeqa
where  $\epsilon_m $ is the Neumann factor which is
equal to 1 for $m=0$ and equal to 2 otherwise.

Since it can be shown that
\beq
\sum_{m=0}^{M} \epsilon_m \cos{m\theta} =
\frac{\sin{(M+\frac{1}{2})\theta}}{\sin{\frac{1}{2}\theta}}
\eeq
The two infinite sums approach $2\pi\delta(\theta)$ and $2\pi\delta(\theta-\pi)$ respectively,
and so the incoming wave comes in from the left, and the outgoing wave goes out
to the right.
The presence of the potential modifies the wave function and in addition to the plane wave,
produces a scattered wave
\beq
\psi  =  e^{ikx} + \frac{e^{ikr}}{\sqrt{r}} f(r\theta)
\eeq
where
\beq
\frac{e^{ikr}}{\sqrt{r}} f(r,\theta)
 = -i \sum_{m=0}^{\infty} \epsilon_m e^{\frac{1}{2}m\pi i - i\delta_m} \sin{\delta_m} 
H_m(kr) \cos{m\theta} \,\,\,\,\, ,
\eeq
$H_m(kr)$ are Hermite polynomials and
\beq
\tan{\delta_m} = \frac{-J_m(ka)}{N_m(ka)} 
\eeq
($N_m(ka)$ are Bessel functions of the second kind).
For large values of $r$, the wave function can be written in a manner similar 
to (\ref{eq:inout}), 
except that
the outgoing wave is modified by the phase shifts $\delta_m$.
\beq
\psi  \simeq \frac{1}{\sqrt{2\pi i k}} 
 i \sum_{m=0}^{\infty} \epsilon_m \cos{m(\theta-\pi)} \frac{e^{-ikr}}{\sqrt{r}}
+ 
\frac{e^{ikr}}{\sqrt{r}} f(r,\theta)   \,\,\, ,
\eeq
where 
\beq
f(r,\theta) \simeq
\frac{1}{\sqrt{2\pi i k}}
\sum_{m=0}^{\infty} \epsilon_m e^{-2i\delta_m(ka)} \cos{m\theta} 
\eeq 
In the limit that $ka \gg m$ the phase shifts can be written as
\beq
\delta_m \simeq ka -\frac{\pi}{2}(m+\frac{1}{2}) \,\,\,\, . \label{eq:shifts}
\eeq 
In the limit of extremely large $a$ (but $r\gg a$), the outgoing waves then
behave as 
\beq
f(r,\theta)\simeq
\lim_{M\rightarrow\infty}
-i\frac{1}{\sqrt{2\pi i k}}e^{-2ika} 
\frac{\sin{(M+\frac{1}{2})(\theta-\pi)}}{\sin{\frac{1}{2}(\theta-\pi)}}
\eeq
where once again we see that the angular distribution goes as the delta function 
$\delta(\theta-\pi)$.  The disk scatters the plane wave directly back, and 
a sharp shadow is produced.  We see therefore, that in the limit of $ka \gg 1$,
our measurement of coincidence works.

The differential cross section can in general be written as 
\beqa
\sigma & = & |f(\theta)|^2 \nonumber\\
&=& | \sum_{m=0}^{\infty} \epsilon_m e^{-2i\delta_m(ka)} \cos{m\theta} |^2
\label{eq:cross}
\eeqa 
For $ka \gg 1$ (but still finite), (\ref{eq:cross}) can be computed using 
our expression for the phase shifts from (\ref{eq:shifts}), and is given
by
\beq
\sigma(\theta)\simeq
\frac{a}{2}\sin{\frac{\theta}{2}} 
+ 
\frac{1}{2\pi k}\cot^2{\frac{\theta}{2}}\sin^2{ka\theta}
\eeq

The first term represents the part of the plane wave which is scattered back,
while the second term is a forward scattered wave which actually interferes with
the plane-wave.  The reason it appears in our expression
for the scattering cross section is because we have written our wave function 
as the sum of a plane-wave and a scattered wave, and so part of the scattered
wave must interfere with the plane-wave to produce the shadow behind the disk.

For $ka \ll m$, the phase shifts look like
\beq
\delta_m(ka) \simeq \frac{\pi m}{(m!)^2} \left( \frac{ka}{2} \right)^{2m} 
\,\,\,\,\,\,\,\,\,\,\, 
m \neq 0 
\eeq
and 
\beq
\tan{\delta_0(ka)} \simeq \frac{-\pi}{2\ln{ka}} 
\eeq

As a result, for $ka\ll 1$,  $\delta_0$ is much greater than all the other
$\delta_m$ and
the outgoing solution is almost a pure isotropic s-wave.

For $ka \ll 1$ the only contribution to (\ref{eq:cross})
comes from $\delta_0$ and
the differential cross section becomes
\beq
\sigma(\theta) \simeq \frac{\pi}{2k\ln^2{ka}} 
\eeq
and is isotropic.  In other words, no shadow is formed at all, and particles are scattered
into classically forbidden regions.  We see therefore, that as long as the s-wave
is dominant, our measurement fails.
The s-wave will cease being dominant when $\delta_0$ is of the same order as
$\delta_1$.  
As can be seen from (\ref{eq:shifts}),
$\delta_1 / \delta_0$ approaches a limiting value of $1$ when a sharp shadow
is produced.  It is only when $\delta_1 / \delta_0 \simeq 1$ that the
cross-section no longer depends on $k$.  This is what we require then, for
the probability of our measurement to succeed independently of
the energy of the incoming particles.
From a plot of $\delta_1 / \delta_0$ we see that this only occurs when $ka
\gg 1$ (Figure 5).  Our condition for an accurate measurement is  therefore
that $a\gg 1/k$.  Since $\delta t_c \simeq a m / k$ we find 
\beq 
\delta t_c \gg 1/E
\eeq

%\begin{figure}
%the begin and end figure is for a figure section, and includes a caption 
%\vspace {3cm}
%\hspace {3.3cm}
%\epsfysize=2.0in
%\epsfbox[-230 132 575 700]{essai2.ps}
% this is another way of doing it
%\epsfbox{phase.ps}
%\caption[]{ 
%$\protect\sqrt{\frac{\Delta}{m}}
%$-arctan(\frac{J_1(ka)}{N_1(ka)})$  vs. $ka$ 
\figlocbig{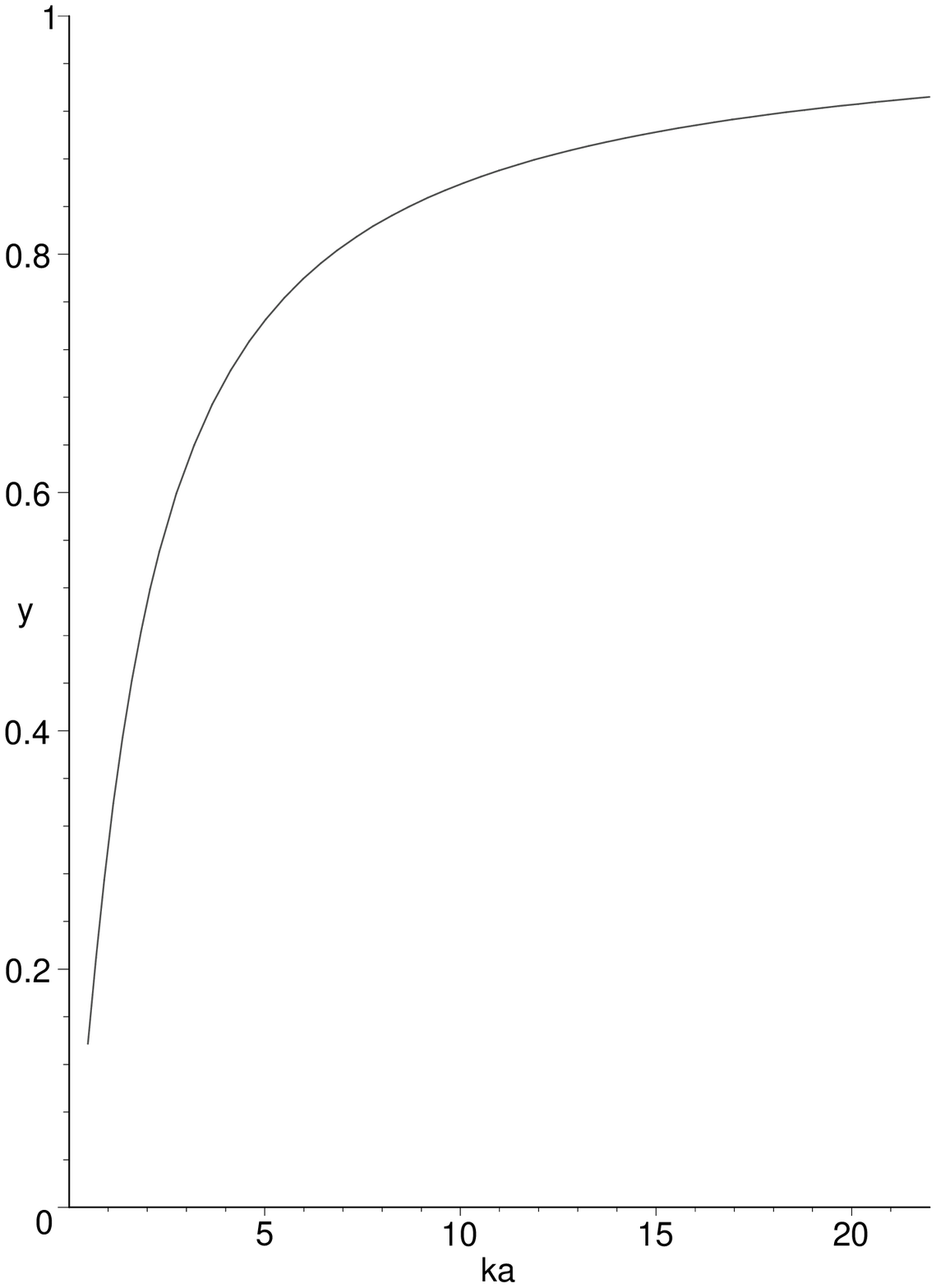}{}{$\delta_1(ka)/\delta_0(ka)$ vs. $ka$}
%\end{figure}

%\section{order of arrival}

\section{Coincident States} \label{or:micro}

%For a conventional observable in quantum mechanics, one can prepare a
%system in an initial state which is an eigenstate of that 
%observable.  If coincidence was a conventional observable,
%one would require that one could prepare two particles in an initial state
We have seen that we can only measure coincidence to an accuracy of
$\delta t_c=1/{\bar E}$.  We shall now show that one cannot prepare
a two particle system in a state
$\psi_c$ which always arrives coincidentally within a time less than 
$\delta t_c$.  In other words, one cannot prepare a system in a state
which arrives coincidentally to greater accuracy than that set by
the limitation on coincidence measurements.

Preparing a state $\psi_c$ corresponds to preparing a single particle 
in two dimensions which always arrives inside a region 
$\delta r=p\delta t_c/m$ 
of the origin.
In other words, suppose we were to set up a detector of size 
$\delta r$ at the origin.  If a state $\psi_c$ exists, then
it would always trigger the detector at some later time.

Our definition of coincidence requires that the state $\psi_c$
not be a state where one particle arrives at a time 
$ t > \delta t_c$ before the other particle.
In other words, if instead, we were to perform a measurement on $\psi_c$
to determine whether particle x
arrived at least $\delta t_c$ before particle y, then we must get 
a negative result for this measurement.

This latter measurement would correspond to the two-dimensional 
experiment of placing a series of
detectors on the positive y-axis, and measuring whether any of
them are triggered by $\psi_c$. If $\psi_c$ is truly a coincident
state, then none of the detectors which are placed at a distance
greater than $y=\delta r$ can be triggered.  One could
even consider a single detector, placed for example, at $(0,\d)$,
and one would require that $\psi_c$ not trigger this detector.

Now consider the following experiment.  We have a particle detector
which is either placed at the origin, or at $(0,\d)$ (we
are not told which).  Then
after a sufficient length of time, we observe whether it has been
triggered.  If we can prepare a coincident state $\psi_c$, then
it will always trigger the detector when the detector is at
the origin, but never trigger the detector when the detector is
at $(0,\d)$.  This will allow us to determine whether the detector
was placed at the origin, or at $(0,\d)$.  For example, if we use the
detectors described in Section \ref{or:coincidence} (namely, just a scattering
potential), then some of the time, the particle will be scattered, and some
of the time it won't be, and if it is scattered, we can conclude
that the potential was centered around the origin rather than around
$(0,\d)$.

However, as we know from Heisenberg's {\it gedanken} microscope experiment, a
particle cannot be used to resolve anything greater than it's wavelength.  In
other words $\psi_c$ cannot be used to determine whether the
detector is at the origin, or at 
$(0,\d) $ if $\d<2\pi/k$.  As a result,
$\psi_c$ can only be coincident to a region around the origin of radius
less than $\d$ or, coincident within a time $\delta t_c \sim 1/E$.

\section{Conclusion}

The notion that events proceed in a well defined sequence is unquestionable
in classical mechanics.  Events occur one after the other, and our knowledge
concerning the events at one time allows us to predict what will occur at
another time.   One can
unambiguously determine whether events lie in the past or future of other
events.  Given two events, $A$ and $B$, one can compute which event occurred
first.  It may be, that event $A$ causes event $B$, in which case, event $A$
must have preceded event $B$.

However, in quantum mechanics the situation is different.  We have argued that
we cannot measure the order of arrival for two free particles if they arrive
within a time of $1/{\bar E}$ of each other, where ${\bar E}$ is their typical
total kinetic energy.  If we try to measure whether they arrive within a time
$\delta t_c$ of each other, then our measurement fails unless we have at least
$\delta t_c > 1/{\bar E}$. Furthermore, we cannot construct a two particle state
where both particles arrive to a certain point within a time of $1/{\bar E}$ of
each other.

Interestingly, this inaccuracy limitation is weaker than what would be
obtained if one tried to measure the time-of-arrival of each particle
separately.

It may be interesting to consider the situation where we have
an event B which {\em must} be preceded by an event A.  For example,
B could be caused by A, or the dynamics could be such that B can
only occur when the system is in the state A.  One can then attempt
to force B to occur as close to the occurrence of event A as possible.
A related   problem has been studied in connection
to the maximum speed of dynamical systems such as quantum computers
\cite{dynamical}  and it was found that one cannot force the system to
evolve at a rate greater than $1/\bar E$ (where $\bar E$
is the average energy), rather than $1/dE$ (where $dE$ is
the uncertainty in the energy).
However since this result
concerns only the free evolution of the system between states,
it is not clear a priori that it is indeed related to the $1/\bar E$
restriction found in the present case where
the measurement interaction disturbs the system.

\vspace{1in}

\noindent{\bf Acknowledgments:}
J.O would like to thank Yakir Aharonov and and Mark Halpern for valuable
discussion.    W.G.U. acknowledges the CIAR and NSERC for support during the
completion of this work.  J.O. also acknowledges NSERC for their support.
B.R. acknowledges the support from grant 471/98 of the Israel Science
Foundation, established by the Israel Academy of Sciences and Humanities.


\vspace{2cm}

\begin{thebibliography}{9}

\bibitem{pauli}
W. Pauli, {\it Die allgemeinen Prinzipien der Wellenmechanik},
in {\it Handbook of physics},
eds. H. Geiger and K. Schell, Vol. 24 Part 1,
(Berlin, Springer Verlag), 1958.

\bibitem{allcock}
G.R. Allcock, Ann Phys {\bf 53} (1969), 253

%\bibitem{peres}
%A. Peres, Am. J. Phys., {\bf 48}, 552 (1980).

\bibitem{muga}
For a review of developments on the arrival time problem see for example
J. G. Muga and C. R. Leavens, Phys. Rep. 338, (2000), 353. or
J.G. Muga, R. Sala, J.P. Palao (quant-ph/9801043),
Superlattice Microst 23: (3-4) 833-842 1998.
\bibitem{aharonov}
Y. Aharonov, J. Oppenheim, S. Popescu, B. Reznik, W.G. Unruh,
(quant-ph/9709031), Phys. Rev. A {\bf 57}, 4130 (1998)
\bibitem{rovelli}
N. Grot, C. Rovelli, R. S. Tate, Phys. Rev. {\bf A54}, 4676 (1996),
quant-ph/9603021.
\bibitem{circ}
The problem of time-of-arrival has also been discussed in the context of
POVM's (e.g. P. Busch, M. Grabowski, P.J. Lahti, Phys. Lett. A 191
(1994) 357). Distributional approaches e.g. J. Kijowski, Rep. Math. Phys. 6, 361
(1974); A. Baute; I. Egusquiza, J. Muga PRA 64, 012501 (2001);
Or within Bohmian mechanics -
C.R. Leavens, Phys. Rev. A 58 (1998) 840;
M. Daumer, in: J.T. Cushing, A. Fine, S. Goldstein (Eds.), Bohmian Mechanics and Quantum Theory:
An Appraisal, Kluwer, Dordrecht, 1996; The interested reader is referred
to reference [3] for a review of the various approaches.
\bibitem{tmeas}
J. Oppenheim, B. Reznik, W.G. Unruh,
quant-ph/9805064, Found. Phys.  .This work also contains arguments
on why time-of-arrival distributions can be problematic, even in the classical
case.

\bibitem{reedsimon}
M. Reed and B. Simon, Methods of Modern Mathematical Physics II: Fourier
Analysis, Self-Adjointness,   New York, Academic Press, 1975.



\bibitem{oppenheim}
J. Oppenheim, B. Reznik, W.G. Unruh,Phys. Rev. A, {\bf 59} 1804 (1999),
quant-ph/9801034; see also
Muga, Leavens and Palao, Phys. Rev. A 58 (1998) 4336.
\bibitem{bennicasher}
For a discussion of this, see A. Casher, B. Reznik "Back-Reaction of Clocks and
Limitations on Observability in Closed Systems", quant-ph/9909010, to appear
in Phys. Rev. A 

\bibitem{heisenberg}
W. Heisenberg, Zeitschrift f\"{u}r Physik, 43, 172-198 (1927), reprinted in {\it
Quantum Theory and Measurement}, J. A. Wheeler and W.H. Zurek, eds (Princeton Univ.
Press, 1983)

%\bibitem{morseandf}
%A. Sommerfeld, {\em Optics}, Lectures on Theoretical Physics, Vol. 4,
%New York, Academic Press 1954.

\bibitem{morseandf}
See for example, Morse and Feshbach, Methods of Theoretical Physics,
McGraw-Hill Book company, New York, 1953 or the rigorous
M. Reed and B. Simon, Methods of Modern Mathematical Physics III: Scattering Theory,   New York, Academic Press, 1975.
 

\bibitem{dynamical}
N. Margolus, L. Levitin, Physica D120 (1998) 188-195, quant-ph/9710043
\end{thebibliography}
\end{document}